\documentclass[fleqn,usenatbib]{mnras}

\usepackage{newtxtext,newtxmath}

\usepackage[T1]{fontenc}

\DeclareRobustCommand{\VAN}[3]{#2}
\let\VANthebibliography\thebibliography
\def\thebibliography{\DeclareRobustCommand{\VAN}[3]{##3}\VANthebibliography}

\usepackage{graphicx}	
\usepackage{amsmath}	
\usepackage{mathtools, cuted}
\usepackage{bm}

\newcommand{\bin}{\mathrm{bin}}

\title[GW searches by missions to Uranus and Neptune]{Searching for gravitational waves via Doppler tracking by future missions to  Uranus and Neptune}

\author[Soyuer et al.]{
Deniz Soyuer,$^{1}$\thanks{E-mail: deniz.soyuer@uzh.ch}
Lorenz Zwick,$^{1}$
Daniel J. D'Orazio$^{2}$
and Prasenjit Saha$^{3}$
\\
$^{1}$Center for Theoretical Astrophysics and Cosmology, Institute for Computational Science, University of Zurich, Winterthurerstrasse 190, CH-8057 Zurich,
\\
Switzerland \\
$^{2}$Niels Bohr International Academy, Niels Bohr Institute, Blegdamsvej 17, 2100 Copenhagen, Denmark\\
$^{3}$Physik-Institut, Universit{\"a}t Z{\"u}rich, Winterthurerstrasse 190, CH-8057 Z{\"u}rich, Switzerland\\
}

\date{Accepted 2021 March 10. Received 2021 March 03; in original form 2021 January 28}

\pubyear{2021}

\begin{document}
\label{firstpage}
\pagerange{\pageref{firstpage}--\pageref{lastpage}}
\maketitle
\begin{abstract}

The past year has seen numerous publications underlining the importance of a space mission to the ice giants in the upcoming decade. 
Proposed mission plans involve a $\sim$10 year cruise time to the ice giants. This cruise time can be utilized to search for low-frequency gravitational waves (GWs) by observing the Doppler shift caused by them in the Earth--spacecraft radio link. We calculate the sensitivity of prospective ice giant missions to GWs.
Then, adopting a steady-state black hole binary population, we derive a conservative estimate for the detection rate of extreme mass ratio inspirals (EMRIs), supermassive-- (SMBH) and stellar mass binary black hole (sBBH)  mergers. We link the SMBH population to the fraction of quasars $f_\bin$  resulting from galaxy mergers that pair SMBHs to a binary.
For a total of ten 40-day observations during the cruise of a single spacecraft, $\mathcal{O}(f_\bin)\sim0.5$ detections of SMBH mergers are likely, if  Allan deviation of Cassini-era noise is improved by $\sim 10^2$ in the $10^{-5}-10^{-3}$ Hz range.  For EMRIs the number of detections lies  between $\mathcal{O}(0.1) - \mathcal{O}(100)$. Furthermore, ice giant missions combined with the Laser Interferometer Space Antenna (LISA) would improve the localisation by an order of magnitude compared to LISA by itself.
\end{abstract}




\section{Introduction}
Uranus and Neptune are the outermost planets in our solar system, orbiting at roughly 20 and 30 AU from the Sun, respectively. Not surprisingly, they are the least explored planets in the system, having been visited only once by the \textit{Voyager II} spacecraft in the late 1980s. The past year has seen numerous white papers calling for missions to the ice giants \citep{whitepaper4,abigail,whitepaper,whitepaper3, dahl}.
The scientific potential of possible missions, along with various mission designs have also been extensively discussed in \citet{hofstadter, fletcher2, fletcher,helled, quest, simon2020, kollmann}. A consensus is being reached in the community regarding the launch date of a possible ice giants mission that would maximize the payload and consequently have a high science yield. The time frame is reported to be around 2029-2030 for Neptune and early 2030s for Uranus, especially if a Jupiter Gravity Assist (JGA) will be used to reach the ice giants \citep{hofstadter}.

Considering that they will spend most of their time in interplanetary space (rather than orbiting the planets they are destined for), the science potential of such mission configurations is limited to a fraction of their lifetime. 
However, if the transponders on the spacecraft were used to detect GWs via Doppler tracking during the cruise phase, these missions would also offer a unique (and \textit{cheap}) opportunity to look for low-frequency GWs.

GWs passing through the space between the transponder and the transmitter/receiver at Earth cause variations in the light travel time between the two, which  correspond to a Doppler shift in the frequency of the transmitted/received signal $\Delta \nu / \nu_0$, where $\nu_0$ is the signal carrier frequency.
Analysis of the resulting Doppler shift allows to reconstruct the strain due to GWs between Earth and the spacecraft, making the Earth-satellite system an arm of a GW observatory.
We expand on the details later and note that
this process is elegantly described in \citet{armstrong} and the references therein.

This use of Doppler tracking with interplanetary spacecraft was previously suggested for many space missions; most notably in the \textit{Pioneer 11} data analysis \citet{pioneer11}, the \textit{Galileo}--\textit{Ulysses}--\textit{Mars} \textit{Observer} coincidence experiment \citep{galileo, ulysses}, and the \textit{Cassini} \citep{cassini92, godtierpaper} mission. 
However, there were  no significant GW candidates  due to insufficient signal levels  \citep{armstrong}. Nevertheless, with increasing capabilities to combat detection noise, we suggest Doppler tracking could be a cheap and efficient way to do science during the cruise phase of prospective ice giant missions.


\section{mission plan} 
\label{sec:missionplan}
There are many proposed designs  for a possible ice giant mission.
A notable one consists of a spacecraft separating into two just before engaging in a JGA, after which each payload travels toward their destined planets (see Fig. \ref{fig:SS}). The mission timeline is projected as:
\begin{itemize}
    \item \textbf{Feb. 2031:} Space Launch System (SLS) departure from Earth.
    \item \textbf{Dec. 2032:} Separation of the spacecraft and subsequent JGA.
    \item \textbf{Apr. 2042:} Arrival of the first spacecraft  at Uranus.
    \item \textbf{Sep. 2044:} Arrival of the second spacecraft at Neptune.
\end{itemize}
We approximate a trajectory for both spacecraft with above timestamps, assuming they leave Jupiter's sphere of influence soon after the JGA. Meaning, their trajectories evolve under solar gravity with the initial asymptotic velocities acquired after the JGA. Fig. \ref{fig:SS} shows orbital positions of the planets and spacecraft trajectories after the JGA, the Earth--spacecraft distance, and the angle subtended by both spacecraft from the Earth.
Although, knowing the exact trajectory is crucial for the actual measurement, it is not important for our sensitivity curve estimation, which will remain qualitatively unaffected by the details of the mission. Both spacecraft enter near-conjunction $\varepsilon > 150^\circ$ for $\sim$2 months at a time, for a total of 9 times for the Uranus spacecraft and 11 times for the Neptune one.
As noted in \citet{armstrong}, the elongation $\varepsilon$ is directly related to  plasma scintillation noise due to  solar winds and irregularities in Earth's ionosphere, where angles $\varepsilon < 150^\circ$ are not ideal for observations.  
For the purposes of our SNR calculations, we take a single Earth-spacecraft system that can collect data for ten 40-day observations, evenly spread within 10 years cruising time between 8 and 30 AU. 
\vspace{-0.7cm}
\section{Methods}
\label{sec:methods}
\subsection{GW response of a Doppler tracking system}
Following \citet{armstrong}, we define the fractional frequency fluctuation of a two-way Doppler system, with  monochromatic carrier frequency $\nu_0$, as
$y_2(t) = {\Delta \nu}/{\nu_0}$, 
where $T_2$ is the two-way light time between Earth and the spacecraft, and $\Delta \nu = \nu(t-T_2) - \nu(t)$.
Then, the frequency fluctuation due to a passing GW can be expressed as
\begin{equation}
    y_2^{\scriptscriptstyle\mathrm{GW}}(t) = \frac{\mu -1}{2} \bar{\Psi}(t) - \mu \bar{\Psi}\left(t - \frac{\mu + 1}{2} T_2 \right) + \frac{\mu+1}{2} \bar{\Psi}(t - T_2),
    \label{eq:y2}
\end{equation}
where $\mu = \hat{\bm{k}} \cdot \hat{\bm{n}}$ is the projection of the unit wavevector $\hat{\bm{k}}$ of the wave onto the unit vector connecting the Earth and the spacecraft $\hat{\bm{n}}$, and $\bar{\Psi}$ is the projection of the GW amplitude
onto the Doppler link \citep{psi}. The latter is given by $\bar{\Psi}(t) = (\hat{\bm{n}} \cdot \mathrm{\bf{h}}(t) \cdot \hat{\bm{n}})/(1 - \mu^2)$,
where $\mathrm{\bf{h}}(t) = \mathrm{\bf{h}}_+(t) \, \mathrm{\bf{e}}_+ + \,\mathrm{\bf{h}}_\times(t) \, \mathrm{\bf{e}}_\times$ is the GW amplitude (i.e. the strain) and $\mathrm{\bf{e}}_{+,\times}$ are the usual "plus" and the "cross" polarization states of a transverse, traceless plane GW. 

\begin{figure}
    \centering
    \includegraphics[width =\columnwidth]{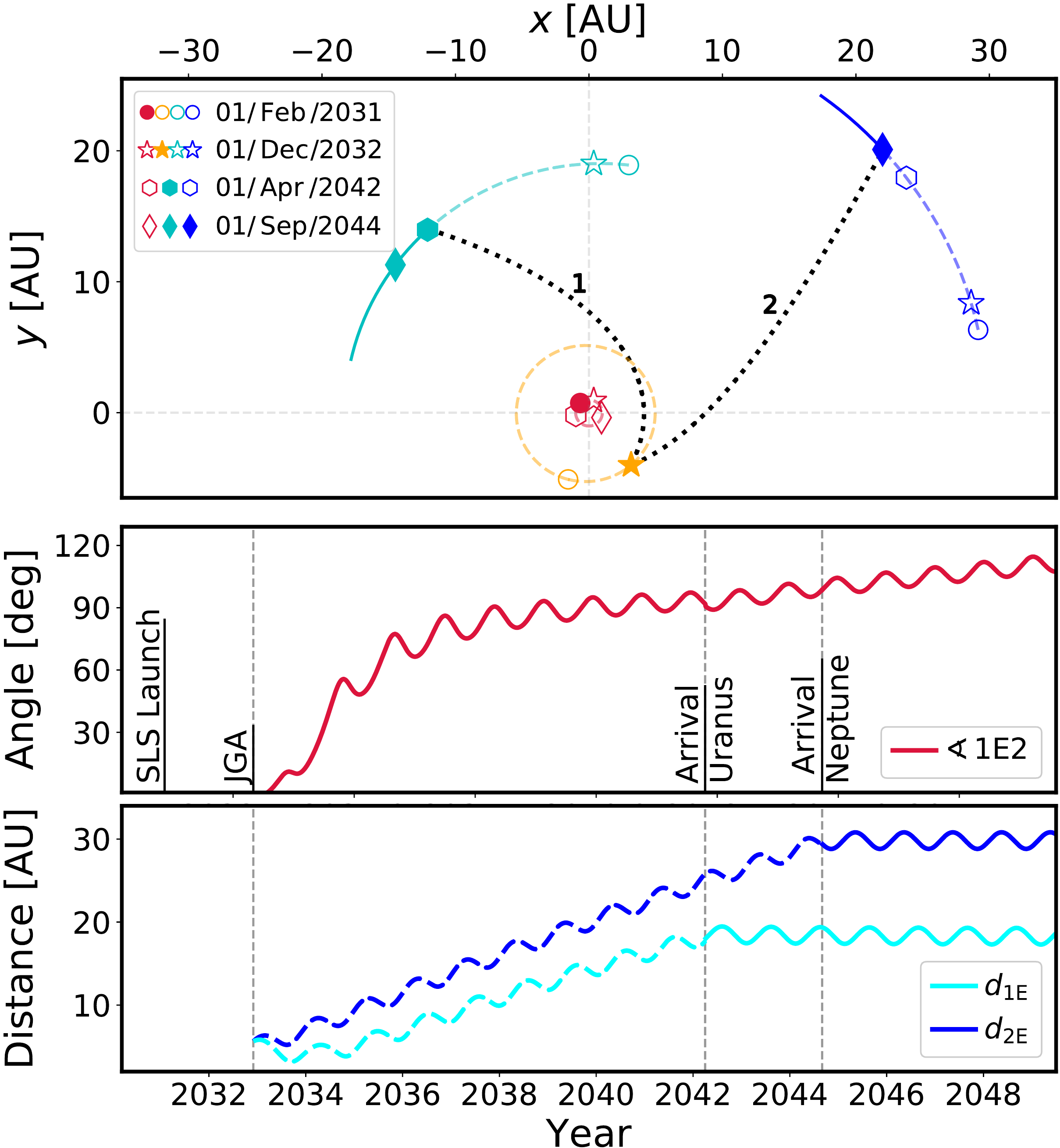}
    \caption{Top panel: Orbital positions of Earth (red), Jupiter (orange), Uranus (cyan) and Neptune (blue) centered around the Sun, plotted until 01/Jan/2050. Markers  represent different timestamps, where full ones indicate the presence of a spacecraft. Location data have been acquired from the JPL HORIZONS System using the \texttt{Astroquery} tool \citet{astroquery}. Mission details given in: {\url{https://github.com/ice-giants/papers/raw/master/presentation/IGs2020_missiondesign_elliott.pdf}}. Center panel: Angle between the Uranus spacecraft (1), Earth (E) and the Neptune spacecraft (2). Relevant timestamps of the mission are  shown with vertical  lines.
    Bottom panel: The distances between spacecraft and the Earth.
    }
    \label{fig:SS}
\end{figure}
\begin{figure}
    \centering
    \includegraphics[width  =\columnwidth]{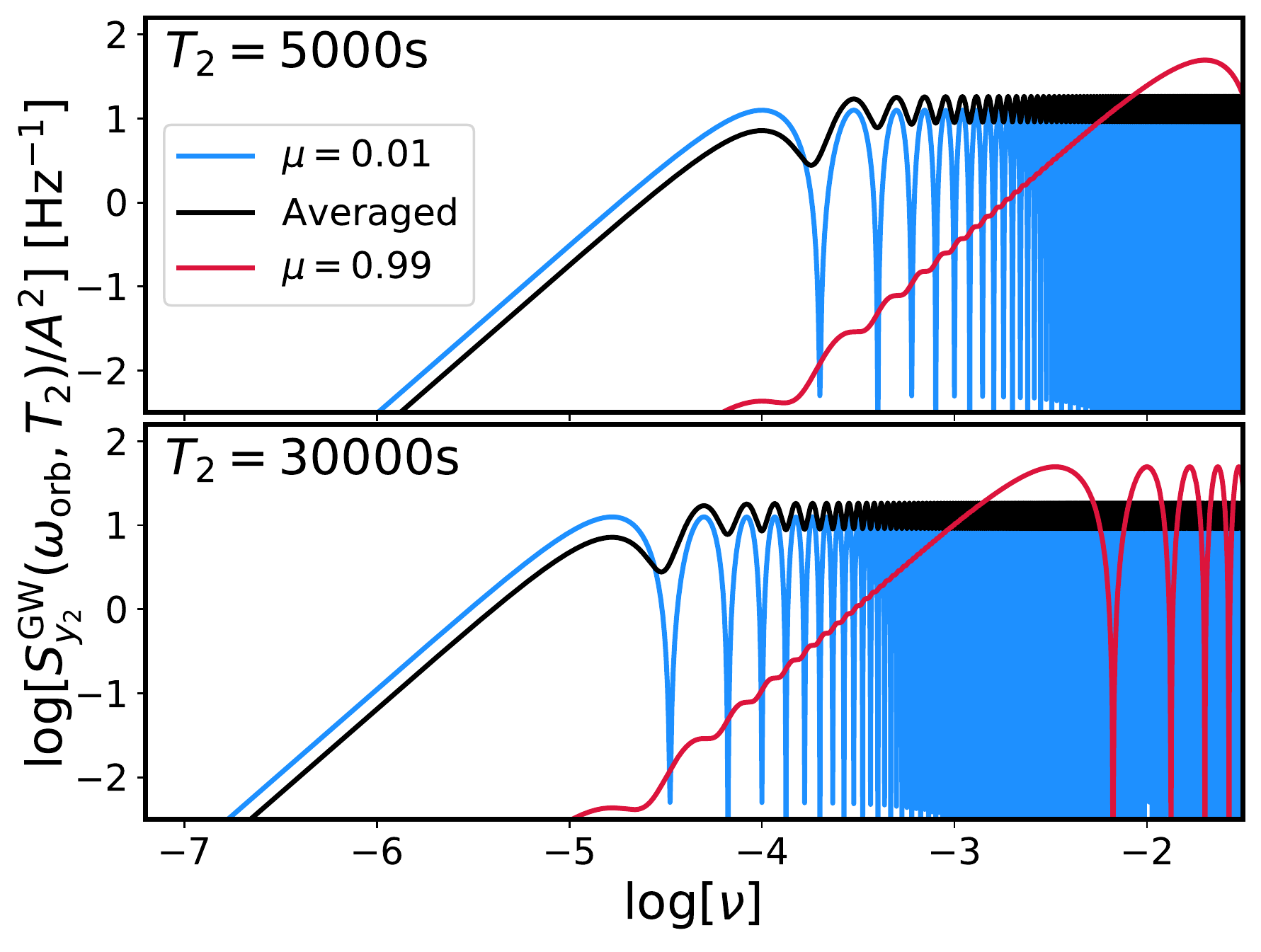}
    \vspace{-0.4cm}
    \caption{The power response of the frequency fluctuations for a GW with $\mu = 0.01$ (blue), $\mu = 0.99$ (red) and sky-averaged (black), for a two-way light time $T_2$ of 5000s ($\approx 5$AU, top panel) and 30000s ($\approx 30$AU, bottom panel).
    }
    \label{fig:power}
    \vspace{-0.2cm}
\end{figure}

Compact binary systems are the most promising sources of GWs, as they produce sinusoidal waves with the form 
\begin{equation}
    \mathrm{\bf{h}}_+(t)\propto A \cos{2\omega_\mathrm{orb} t} \,\mathrm{\bf{e}}_+, \,\,\,\,\,\,\, \mathrm{\bf{h}}_\times(t) \propto 4 A  \sin{2\omega_\mathrm{orb} t}\, \mathrm{\bf{e}}_\times,
\end{equation} 
where $\omega_\mathrm{orb}$ is the angular frequency of the binary system and $A$ the monochromatic instantaneous amplitude of the passing GW \citep{moore}. The spectral power response $S_{y_2}^{\scriptscriptstyle\mathrm{GW}}$ of this fluctuation is a measure of the noiseless sensitivity of a Doppler system to a GW of a particular frequency, and can be calculated analytically for a sinusoidal source by taking the Fourier transform of the fluctuation $\tilde{y}_2 =  \mathcal{F}[y_2^{\scriptscriptstyle\mathrm{GW}}]$ and reading the coefficient of the Dirac delta function
\begin{equation}
    \left|\tilde{y}_2(x, \omega_\mathrm{orb}, T_2)\right|^2 =S_{y_2}^{\scriptscriptstyle\mathrm{GW}}(\omega_\mathrm{orb}, T_2) \, \delta(x - 2\omega_\mathrm{orb}).
\end{equation}
Fig. \ref{fig:power} shows $S_{y_2}^{\scriptscriptstyle\mathrm{GW}}(\omega_{\rm{orb}}, T_2)$ of a source located in two extreme directions (blue and red) and the sky-averaged response (black), for two different light travel times. 
\vspace{-0.2cm}
\subsection{Sensitivity of an ice giant mission}
The total signal produced by a monochromatic GW source does not only depend on its instantaneous gravitational luminosity, but also on the number of GW cycles passing through the detector. The frequency and amplitude of the GW  evolve as the binary shrinks over many orbits.
The characteristic strain  takes into account the cycles that a binary completes in the proximity of a given GW frequency $\nu$
\begin{equation}
h_c(\nu) = \sqrt{\nu ^2\dot{\nu}^{-1}}h_0.
\label{Eq:hc1}
\end{equation}
For the purposes of this letter it is sufficient to use phenomenological waveform amplitudes like the ones in \citet{2007CQGra..24S.689A}, which correctly model the frequency scaling of the inspiral, merger and ringdown phases. 
The characteristic strain amplitudes read
\begin{align}
 h_{\rm c}(\nu) &= \sqrt{\frac{24}{5}} \frac{\left(G \mathcal{M}/c^3\right)^{5/6} \nu_0^{7/6} \nu}{\pi^{2/3}(D_{\rm L}/c)} \begin{cases} 
      \left({\nu_0}/{\nu} \right)^{7/6} & \nu\leq \nu_0 \\
      \left({\nu_0}/{\nu} \right)^{2/3} & \nu_0 < \nu\leq \nu_1 \\
      w \mathcal{L}(\nu) & \nu_1 < \nu \leq \nu_3
   \end{cases}
\end{align}
where $\mathcal{M} = (M_1M_2)^{3/5} / (M_{\mathrm{tot}})^{1/5}$ is the chirp mass and $D_{\rm L}$ the luminosity distance of the binary. $w$ and $\mathcal{L}$ model the decay of the so called "quasi-normal modes"; oscillations of the shape of the event horizon just after  merger, given by
$w \mathcal{L}(\nu) = ({\nu_0/\nu_1})^{2/3}(4(\nu-\nu_1)^2/\nu_2^2 \, -\, 1)^{-1}$.
The frequencies $\nu_0$, $\nu_1$, $\nu_2$ and $\nu_3$ correspond to the merger, ringdown, ringdown decay-width and cut-off frequencies, respectively. Their values are computed as 
$
\nu_{k}=c^3 (a_k \eta^2 + b_k \eta + c_k)/(\pi G M_{\rm{tot}}),
$
where $\eta = (\mathcal{M}/M_{\mathrm{tot}})^{5/3}$ is the symmetric mass ratio and $a_k$, $b_k$ and $c_k$ are conveniently tabulated in \citet{Robson_2019}.

It is important to note that eq. (\ref{Eq:hc1}) is only valid if the source can complete all of the cycles at a given frequency within one observation time $T_{\rm{obs}}$. The limiting frequency for this condition to be true reads
\begin{equation}
    \nu_{\mathrm{lim}} \simeq
    \left(\frac{c^{15} (M_1+M_2)}{G^5 M_1^3 M_2^{3}T_{\mathrm{obs}}^3}\right)^{1/8},
\end{equation}
For GWs at frequencies $\nu \leq \nu_{\mathrm{lim}}$, the number of cycles in a frequency bin is bounded by the observation time. In this case we have
\begin{equation}
 h_{\rm c}(\nu) = \sqrt{\nu T_{\mathrm{obs}}}  h_0.
 \label{Eq:hc2}
\end{equation}

The last step required to produce a sensitivity curve for the Doppler system is to equate the spectral power response of the signal $S_{y_2}^{\scriptscriptstyle\mathrm{GW}}$ and  that of the noise $S_n$, and subsequently solve for the required instantaneous strain amplitude that leads to an SNR of 1:
\begin{equation}
    h_{n}(\nu,T_{\mathrm{obs}}, T_2) = \left(\frac{A^2 S_{n}(\nu,T_{\mathrm{obs}}, T_2)}{\omega_{\rm{\scriptscriptstyle{GW}}} S_{y_2}^{\scriptscriptstyle\mathrm{GW}}(\omega_{\rm{orb}}, T_2) \, T_{\mathrm{obs}}}\right)^{1/2},
    \label{Eq:hn}
\end{equation}
where $S_n$ is for 40 days observation time, same as that calculated by \citet{godtierpaper}. If the characteristic strain of a source $h_c$ is equal to $h_n$, then that source would have an accumulated SNR = 1.

A compilation of  noise sources of the   Cassini-era observations can be seen in table 2 of \citet{armstrong}.
We follow \cite{godtierpaper} to model  $S_{n}(\nu)$ as a triple  power-law, corresponding to three frequency regimes that are dominated by  different noise sources 
\begin{equation}
    S_n(\nu) = S_0 \sum_{i = 1}^3 \left(\frac{\nu}{\nu_i}\right)^{\alpha_i},
    \label{eq:noise}
\end{equation}
where $\nu_1 < \nu_2 < \nu_3$, and $S_0$ is related to the Allan deviation $\sigma_y$ via
$S_0 = T_2 \, \sigma_y^2$. 
For Cassini, the powers read $\alpha_1=-2$, $\alpha_2=-1/2$ and $\alpha_3=2$, where the last two model the frequency propagation noise and the onset of thermal noise, respectively. The steep scaling of the low-end does not have a physical origin, but rather it is a conservative estimate due to the lack of sophisticated data analysis at that regime.

\begin{figure*}
    \centering
    \includegraphics[width =2\columnwidth]{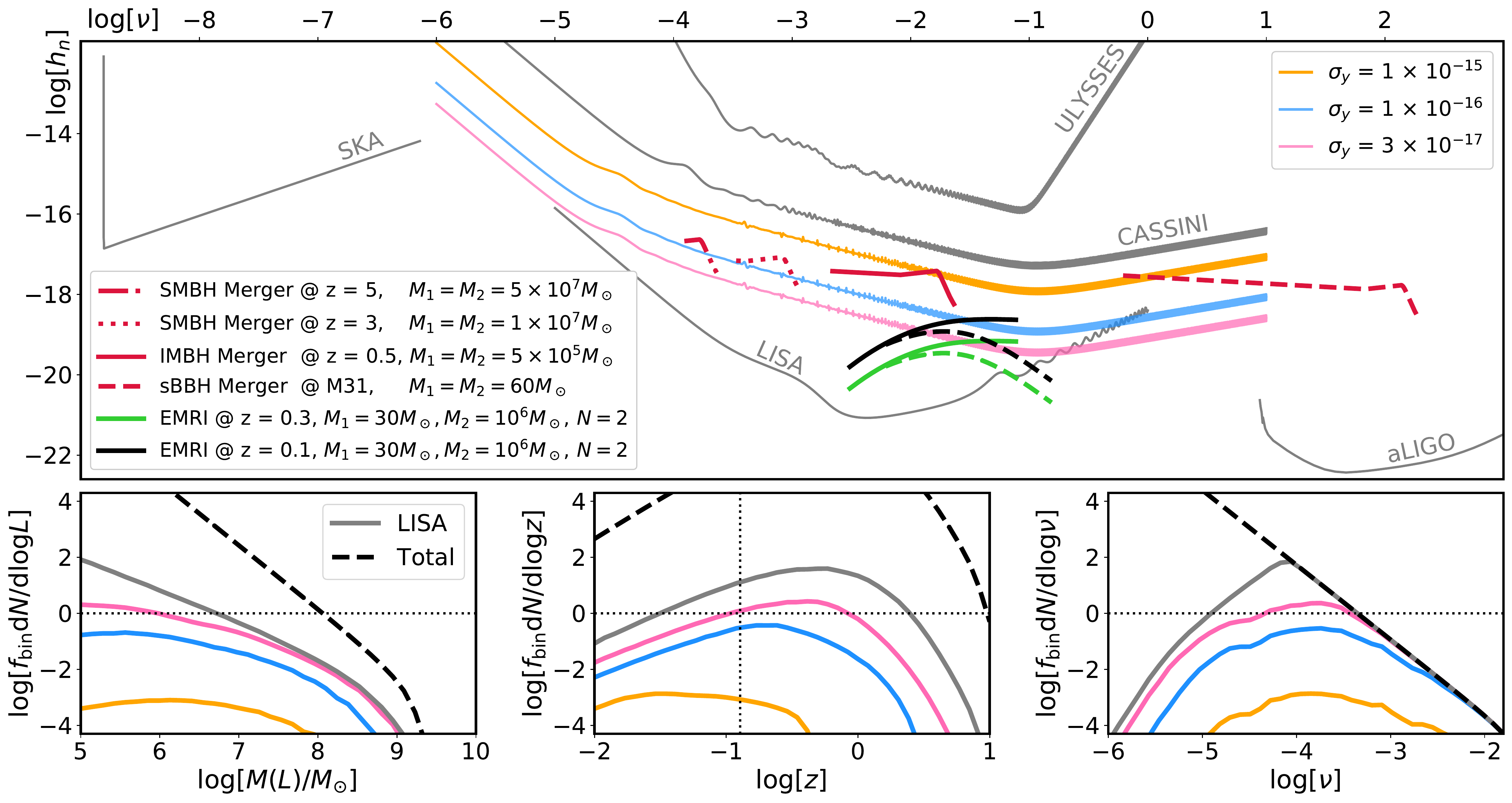}
    \caption{Top panel: Sensitivity of various experiments expressed in $\log h_n$ vs. $\log \nu$ (\textit{Ulysses} and \textit{Cassini} with data from \citet{godtierpaper}, LISA curve from \citet{Robson_2019}, aLIGO and the SKA-era PTA from \citet{moore}). Orange, blue and pink curves represent  sensitivities of the ice giants mission, 
    with an Allan deviation improvement of 3, 30 and 100 times compared to the Cassini-era, respectively.
    We sample 10, 40-day observations between $T_2$ =  8000s and 30000s.
    The red curves correspond to various equal mass GW sources. The green and black curves are the $N=2$ (solid) and $N=4$ (dashed) harmonics of a eccentric EMRIs at $z=0.3, 0.1$. Bottom panels: Number of detectable SBHBs per log mass, redshift, and observed GW frequency. Dashed lines represent all SBHBs predicted by our model, while the solid lines represent the number of SBHBs detectable with the specified detectors $N_{\rm{SBHB}}$ via eq. (\ref{Eq:RateInt}), with a SNR threshold of $\rho_c=1$ (partially justified due to undercounting by $\mathcal{D}$). A luminosity distance of 600 Mpc is denoted in the redshift plot for clarity.}
    \label{fig:SC}
    \vspace{-0.2cm}
\end{figure*}

Fig. \ref{fig:SC} shows the sky-averaged sensitivities of various experiments, as well as those of the prospective ice giant missions. We focus on three cases, where the total Allan deviation is improved by a factor of 3, 30 and 100 with respect to Cassini-era values. As mentioned earlier, we consider ten 40-day observations over 10 years between 8 and 30 AU, where we add the SNR of individual observations in quadrature.
While the noise scaling at low and high-end is uncertain, the cutoff frequencies at $\nu_{\rm{min}}$ $ =\! 2/T_{\rm{obs}}$$\sim$$10^{-6.2}$Hz  and $\nu_{\rm{max}}\!  = \! 2/t_{\mathrm{res}}$$\sim$$10$Hz  are set by   observation and  resolution time, respectively. Hence, sensitivity of an ice giant Doppler tracking GW detector spans from the high-end of  currently operating Pulsar Timing Arrays \citep[PTAs][]{PTAastro:2019}, through the LISA band \citep{LISA:2017}, reaching the low-end of  ground based detectors \citep[\textit{e.g.}, aLIGO][]{aligo}.

\vspace{-0.5cm}
\section{Event rate estimates}
\label{sec:rate}
Expected sources of GWs in this range are the late inspiral and mergers of SBHBs \citep{kleinrev} at the low-end, to the EMRIs \citep[][]{emri} at mid-frequencies, to pre-merger inspirals of  sBBHs that will merge weeks to years later in the LIGO band \citep{SesanaMB:2016,GerosaMB:2019}. Fig. \ref{fig:SC} shows the characteristic strains of various merger events with generic masses and distances.

To estimate event rates for each type of source, we consider binaries with  mass ratio $q=M_2/M_1\leq1$, and zero orbital eccentricity such that they emit GWs at twice the orbital frequency $\nu=2\nu_{\rm{orb}}$. We assume that the number of binaries per orbital frequency is governed by a steady-state continuity equation with orbital frequency evolution dominated by GW emission, $\dot{\nu}_{\scriptscriptstyle \mathrm GW}$. The first assumption is valid when the binaries are formed at $\nu_{\rm{orb}}$ lower than the detection frequencies \citep{ChristianLoeb:2017}, and the second for binaries at late inspiral, both  well motivated here.
Thus, the number density of binaries is
\begin{equation}
   n_{\bin} = \int{ \frac{ \mathrm{d} \mathcal{R}}{\mathrm{d}M} \frac{\mathrm{d}\nu}{ \dot{\nu}_{\scriptscriptstyle \mathrm GW}} \mathrm{d}M},
\end{equation}
where $\mathrm{d} \mathcal{R}/\mathrm{d}M$ is the volumetric merger rate per total binary mass for a specified population, and where we have further assumed that the binary mass and mass ratios are fixed over the observation time.

The total number of these binaries that a detector with sensitivity $S_n$ may detect above a threshold SNR $\rho_c$, is found by integrating over volume $V$ and including only sources above the detection threshold:
\begin{align}
    \label{Eq:RateInt}
   N_{\bin} &= \int\limits^z_0 \int\limits^{\infty}_{0} \int\limits^{\nu}_{\nu_{\rm{ISCO}}}{ \frac{\mathrm{d}V}{ \mathrm{d}z} \frac{ \mathrm{d} \mathcal{R}}{\mathrm{d}M} \mathcal{D}\left( M, q, \nu, z \right) \frac{\mathrm{d}\nu}{ \dot{\nu}_{\scriptscriptstyle \mathrm GW} } \mathrm{d}M \mathrm{d}z}, \\
    &\mathrm{with}\,\,\,\,\,\,\mathcal{D}\left( M, q, \nu, z\right) \equiv  \mathcal{H}\left[ h_c(M, q, \nu, z) -  \rho_c h_{n}(\nu)\right], 
\end{align}
where $\mathrm{d}V/\mathrm{d}z$ is the angle integrated cosmological volume element in a flat universe \citep{HoggCosmoDist:1999} and $\mathcal{H}$ is the Heaviside function. We have assumed the population can be modeled with a single representative mass ratio $q$, which is a variable of the model. The detection probability $\mathcal{D}$ is built from the $h_c$ of eq.s (\ref{Eq:hc1}) and (\ref{Eq:hc2}) and the the characteristic noise, $h_n$, explained surrounding eq. (\ref{Eq:hn}).
We note that $\mathcal{D}$ undercounts the SNR of a source as it compares $h_c$ and $h_n$ at a single frequency rather than integrating over all source frequencies.


\vspace{-0.2cm}
\subsection{Stellar mass binary black hole mergers}
For sBBHs that will eventually merge in the LIGO band, we can reliably tie the merger rate to the measured LIGO merger rate. As only the most massive sBBHs are expected to be observable at low frequencies and sensitivities of interest here \citep{GerosaMB:2019} we estimate the differential merger rate as a fraction of the LIGO inferred merger rate $f \mathcal{R}_{\rm{LIGO}} \delta(M-M_*)$. 
For binary parameters representative of the massive sBBHs detected by LIGO, $M_*=60 M_\odot$ and $q=1$ \citep{ligocat}, eq. (\ref{Eq:RateInt})
shows that even for a $10^2$ improvement in $\sigma_y$, we would expect $0.02$ detectable binaries with $\rho_c=1$. This is consistent with assuming  $\mathcal{D}\left( M, q, \nu, z \right)$ results in a cut of sources beyond a maximum distance $D_{\rm{max}}$ and minimum frequency $\nu_{\rm{min}}$:
\begin{align}
N_{\rm{sBBH}} &\!\approx\! \frac{4 \pi D^3_{\rm{max}}}{3} \mathcal{R}_{\rm{LIGO}} \frac{8}{3}  \left[\frac{\nu}{ \dot{\nu}_{\scriptscriptstyle \mathrm GW} } \right]_{\nu_{\rm{min}}} \\
&\approx 0.02 \left(\frac{\nu_{\rm{min}}}{10^{-2.0}\mathrm{Hz}}\right)^{-8/3} \!
\left(\frac{D_{\rm{max}}}{20\mathrm{Mpc}}\right)^3\!
\left(\frac{f\mathcal{R}_{\rm{LIGO}}}{10 \mathrm{Gpc}^{-3} \mathrm{yr}^{-1}}\right).
\end{align}
For comparison, with LISA SNR cuts of $\rho_c = 2, 5, 8$, one expects to detect $200$, $14$, and $3$ sBBHs respectively, in line with recent estimates \citep{SesanaMB:2016, GerosaMB:2019}. 
\vspace{-0.3cm}
\subsection{Extreme mass ratio inspirals}
We follow a similar procedure for EMRIs, where instead of a measured merger rate, we rely on a range of rates predicted in the literature for various EMRI production channels \citep[see][]{ChenHan_EMRIs2018}. This varies from $\sim10^{-9}-10^{-6}$/gal/yr. Assuming a characteristic EMRI consisting of a $30M_\odot$ BH inspiraling to a $10^6 M_\odot$ SMBH, we find (see Fig. \ref{fig:SC}), such EMRIs fall above the intermediate (blue)  sensitivity curve for $z\!\lesssim\!0.1$. The total number of galaxies within $z=0.1$ is approximately $10^{-2}\mathrm{gal}/\mathrm{Mpc}^3 \times 4\pi (420\mathrm{Mpc})^3/3 = 3 \times10^6$, so we find that the expected number of detectable EMRIs ranges from $3\times10^{-3}$ to $3$ over the lifetime of the proposed mission. Considering the highest sensitivity (pink) mission, the expected number of  EMRI events approaches $\sim$ $0.08$ to $80$ over the mission lifetime.
\vspace{-0.4cm}
\subsection{Supermassive black hole binary mergers}
To compute a differential merger rate for SBHBs, we assume 
a plausible merger channel, namely 
that a fraction $f_{\bin}$ of  quasars result from galaxy mergers that pair SMBHs into a binary at the nucleus of the newly formed galaxy \citep{HKM:2009, DOrazioLoeb:2019}. In other words,  if $f_\bin=1$, then all quasars counted in the observationally determined quasar luminosty function (QLF) have a binary at some stage in its evolution, (albeit not necessarily at the stage we want for the frequency to be high enough to detect).
The number of SBHBs is traced by the quasar population which is itself traced by the  observationally determined QLF: $\mathrm{d}^2N_Q/(\mathrm{d}L\mathrm{d}V)$. Assuming this fraction of the quasars facilitates a SBHB over the quasar lifetime $\tau_Q$, the differential merger rate for SBHBs reads
\begin{equation}
   \left. \frac{ \mathrm{d} \mathcal{R}}{\mathrm{d}M}\right|_{\mathrm{SBHB}} = \frac{f_{\bin}}{\tau_Q} \frac{\mathrm{d}^2N_Q}{\mathrm{d}L\mathrm{d}V} \frac{\mathrm{d}L}{\mathrm{d}M}
   \label{Eq:dRdMSBHB}
\end{equation}
where we relate the quasar bolometric luminosity $L$ to $M$, assuming  the binary accretes at some fraction of the Eddington rate $f_{\mathrm{Edd}}$,
\begin{equation}
M(L) =  L \sigma_T/(f_{\mathrm{Edd}} 4 \pi G m_p c),
\label{Eq:MofL}
\end{equation}
with $L_{\mathrm{Edd}} = 4 \pi G M m_p c /\sigma_T$ the Eddington luminosity. We assume an average value of $f_{\mathrm{Edd}}=0.1$ for bright quasars \citep{SWM:2013_fEdds} that are traced by the QLF from \citep{HopkinsQLF+2007}. 
We then need to know the distribution of these SBHBs in emitted GW frequency and characteristic strain.
The number of binaries per frequency in a steady-state population, assuming circular orbits, can be estimated by assuming binaries are driven together by GW emission. Under this assumption, the fraction of binaries at frequency $\nu$ is approximately the residence time of the binary $\nu/\dot{\nu}_{\scriptscriptstyle\mathrm{GW}} \propto \nu^{-8/3}$ \citep{Sesana+2005, ChristianLoeb:2017, DOrazioDiStefano:2020} divided by the binary lifetime. As fiducial parameters we choose a quasar lifetime of $\tau_Q=10^7$~yr \citep{PMartini:2004}, and $q=0.3$, typical of major galactic mergers  \citep{Volonteri+2003}.

Using eqs. (\ref{Eq:dRdMSBHB}) and (\ref{Eq:MofL}) in eq. (\ref{Eq:RateInt}), and integrating yields estimates for the detectable SBHB population scalable in terms of the $f_{\rm{bin}}$ and the $\tau_Q$. Bottom panels of Fig. \ref{fig:SC} show our estimates for the corresponding ice giant Doppler ranging missions and for a nominal LISA mission, for $f_{\bin} =1$. We see that our most optimistic  detector could observe of order a few SBHB inspiral/mergers at redshifts of $\sim 0.1-1.0$ and frequencies of $10^{-4.5}-10^{-3.5}$Hz. 
Interestingly, our model predicts that such a detection is most likely for an inspiralling, lighter SBHB at $M\leq 10^6 M_\odot$. This is because the quasar (and hence SMBH) population is largest at these smaller masses.

While our model relies on the uncertain means by which SBHBs are brought together and merge, we note that by including only a merger channel that traces the quasars, we tie our estimate to a relevant observable quantity while also conservatively underestimating high redshift mergers. Our estimates are in approximate agreement with the low redshift predictions from more general SBHB population estimates \citep{WyitheLoeb:2003, Klein+2016}.

\begin{figure}
    \centering
    \includegraphics[width =0.93\columnwidth]{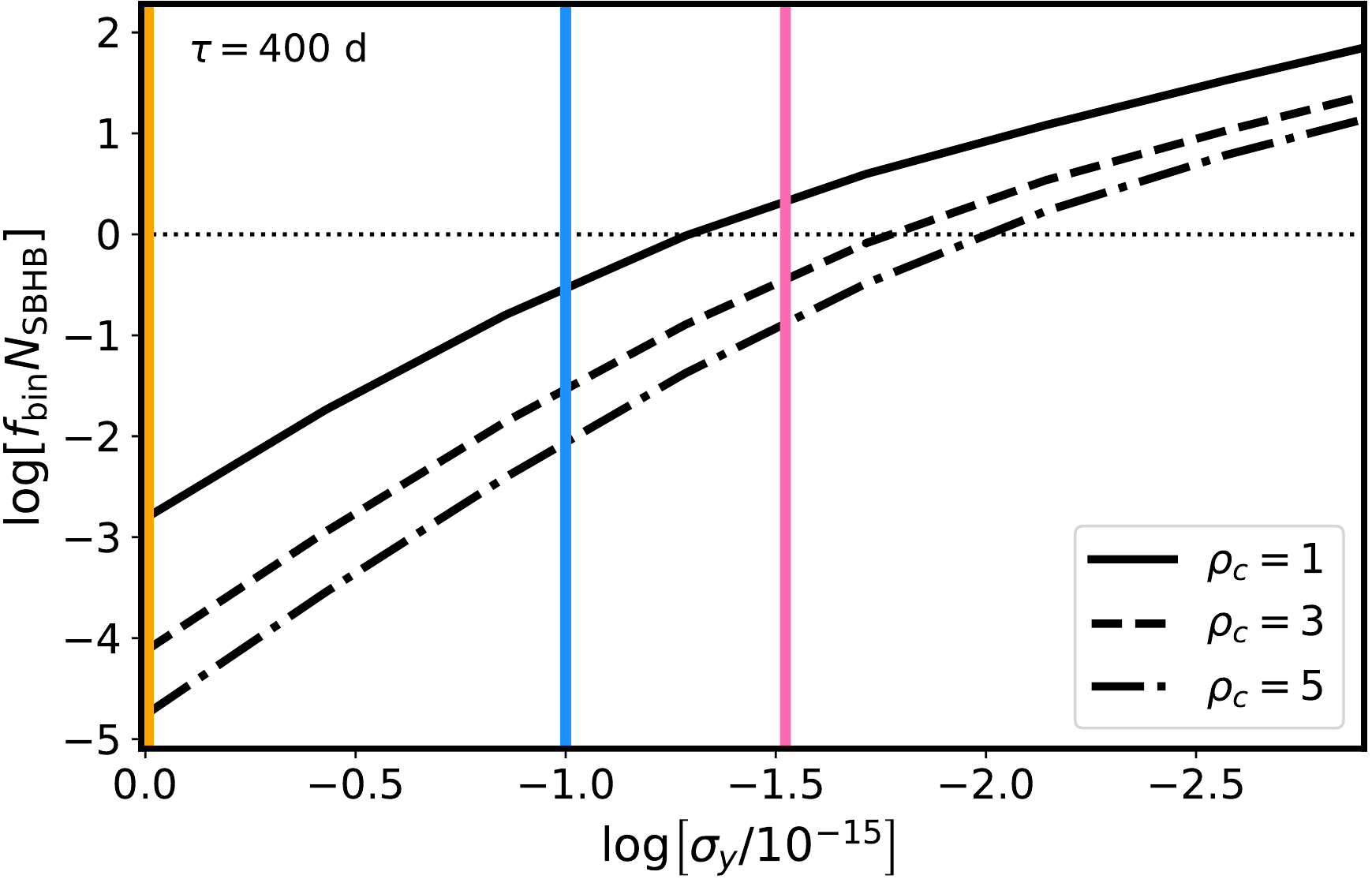}
    \caption{Expected detections of SMBH binary mergers vs. Allan deviation for 400 days of total observation. Black curves represent SNRs of 1,3 and 5. Vertical colored lines are the ice giant missions with Allan deviation improvements of 3, 30 and 100 with respect to the Cassini-era measurements.}
    \label{fig:last}
    \vspace{-0.3cm}
\end{figure}

\vspace{-0.3cm}
\section{Sky Localization of GW sources}
\label{sec:local}
It is a major advantage that ice giant missions would be concurrent with LISA, since both experiments would be likely to detect the same signal with completely independent systems, reducing systematic noise and improving sky localization. The long one-way light travel times $T_1$ of Doppler tracking missions greatly help with the latter, as the localized area on the sky is proportional to 1/$T_1$.
Following
\citet{triang}, the source can be localized on the sky on an area with probability $p$, using 3 detectors as
\begin{equation}
{\rm Area}(p, \theta)/4 \pi \approx  -\ln(1-p)\sigma_x \sigma_y /\cos\theta,
\end{equation}
where $\theta$ is the angle between the normal of the detector plane and the line of sight to the source, and $\sigma_{x,y}$ (not to be confused with Allan deviation) are the relative timing uncertainties along the (x,y) directions (normalized by $T_1$ of the Doppler tracking missions).

Due to the difference in sensitivities between the Doppler tracking missions and LISA, a detection by the former would imply very high SNR for the same detection by the latter. For the Doppler tracking missions, the uncertainty in detection timing lies in resolving the phase of the wave, rather than the time resolution of the Earth--spacecraft link. To simulate this effect, we have run several simple numerical experiments, in which we try to reconstruct the phase of a noisy sinusoidal wave. We find that the uncertainty in the fitted phase scales proportionally to SNR$^{-1} r^{-1/2}$, where $r=(\nu_{\rm \scriptscriptstyle GW}  t_{\rm{res}})^{-1}$ is the amount of data points per period. $\sigma_\phi$ is then conservatively
\begin{equation}
\sigma_\phi \approx   \sqrt{\nu_{\rm \scriptscriptstyle GW} t_{\rm res}} (\nu_{\rm \scriptscriptstyle GW} \times {\rm  SNR})^{-1} \gg \sigma_{\rm \scriptscriptstyle LISA},
\end{equation} 
where the SNR is of the Doppler tracking missions. Note that this timing uncertainty due to $\sigma_{\phi}$ is much greater than the timing uncertainty of LISA, hence the former is the dominating factor in $\sigma_{x,y}$.
In a realistic layout where LISA and two Doppler tracking satellites form a right triangle (LISA at the perpendicular) with catheti  15 and 20 AU, they localizes the source to
\begin{align}
{\rm Area}(p = 90\%, \theta = 60^\circ)/4 \pi &\approx - 2\ln(0.01)\left(\frac{\sigma_\phi}{7500{\rm s}}\right)\left(\frac{\sigma_\phi}{10000{\rm s}}\right) \nonumber\\
&\approx  10^{-9}\nu_{\rm \scriptscriptstyle GW}^{-1}{\rm  SNR}^{-2} \, {\rm s}^{-1}
\end{align}
of the sky  for a median source location at $\cos\theta = 1/2$ and a resolution time of $t_{\rm{res}} =  0.2$ seconds \citep{armstrong}.

\cite{KocsisLocLISA+2007} find that by the time of merger, LISA alone can localize an inspiraling $2\times 10^6 M_{\odot}$ SBHB at $z=1$ to sky fractions of $10^{-4}$ (of order 10 deg$^2$) at 90 per cent confidence. In this frequency range ($\nu_{\scriptscriptstyle \rm GW} \!\sim \!10^{-4}$Hz), the addition of a Doppler tracking experiment could enhance the sky localization by a factor of 10, with a Doppler tracking SNR of 1 and a time resolution $t_{\rm{res}}$ equivalent to that of Cassini   \citep{armstrong}. While our estimate is crude, the order of magnitude suggests that it could greatly enhance LISA science gains by increasing the probability of identifying a host galaxy or discovering an EM counterpart, allowing multi-messenger cosmological, astrophysical and gravitational tests \citep{KocsisLocLISA+2007, local}.

\vspace{-0.1cm}
\section{discussion and conclusion}
\label{sec:disc}
In this letter, we have calculated the feasibility of detecting GWs from BH mergers with a prospective ice giants mission by constructing a Doppler tracking sensitivity curve and implementing a population synthesis model to estimate the merger detection rate.
As shown in Fig. \ref{fig:last}, the Allan deviation, $\sigma_y$,
is a crucial factor in determining whether the  mission will be successful as a GW observatory. Within our conservative merger population model, we find that an improvement in $\sigma_y$ of  2 orders of magnitude over the Cassini parameters is required to have a reasonable chance of detecting a few GW events by SBHBs, and a few to tens of EMRIs, depending on the accuracy of the merger rate estimates. There is also some uncertainty in $f_\bin$, where some estimates put it around $f_\bin \sim 0.25$ \citep{fbin}.

It is hard to predict how much the total Allan deviation is likely to improve over the years, since it depends on several different technologies \citep[as discussed thoroughly in][]{armstrong}. However, a few qualitative arguments suggest that a factor between $10$ to $10^2$ might not be improbable. 
Cassini-era noise is dominated by 3 sources, namely, the antenna mechanical noise, and plasma and tropospheric scintillation noises. The antenna mechanical noise can be attenuated significantly by using a smaller and stiffer antenna in combination with the main dish as demonstrated in \citet{antenna}.

Additionally, noise due to plasma scintillation peaks around $3\times10^{-3}$ Hz and then steadily drops down for lower frequencies \citep{plasma}. Since the main contribution to our detection rate for SMBH mergers comes from $10^{-5} - 10^{-3}$ Hz region, we also expect significantly less contribution to noise in this regime from plasma scintillation.
As noted in \citet{godtierpaper, arms2}, the noise levels for frequencies $\leq 10^{-4}$ Hz has not been extensively studied (hence the uncertain low-frequency scaling in the Cassini noise power law). The tropospheric noise starts to increase with decreasing frequency below $10^{-4}$ Hz and is the dominant noise factor in the region of interest. Water-vapor-radiometer-based corrections to Cassini-era tropospheric noise have improved the Allan deviation by a factor of 2--10 down to $1.5 \times 10^{-15} - 3 \times 10^{-15}$ \citep{armstrong}. Naturally, a significant improvement in tropospheric noise correction is needed to achieve the desired noise levels for detecting SMBH mergers with confidence. One way to reduce the tropospheric noise is using high altitude facilities (on the ground or on balloons) for communication with the spacecraft. Another way of reducing tropospheric noise is using multiple measurement points \citep{tropo}, e.g. via a radio telescope array.
Additionally, with recent advances in optical Doppler orbitography \citep{orbito}, an optical Doppler link (which would have 4 orders of magnitude improvement in atmospheric phase noise compared to typical X-band links) might not be out of the question for the prospective ice giant missions.
Lastly, with the emergence of highly stable optical lattice clocks \citep{clock}, there should be no limitation coming from clock comparisons.

A hypothetical ice giant mission would be scheduled for the 2030s, meaning three decades of technological development from the Cassini-era. Moreover, the expertise gained by the recent success of LIGO and the LISA pathfinder mission is likely to have significant crossover to a Doppler tracking system. 

Reducing Allan deviation of the Doppler tracking system is also crucial for the in situ exploration of ice giants; most importantly,  measuring the Newtonian gravity multipole harmonics $J_n$ of the planets. Precise measurement of these are essential for interior structure modelling, and is the main motivation in improving the Allan deviation of the Doppler link. Additionally, an improvement in Allan deviation would likely enable the detection of relativistic frame-dragging effects due to the planets spin
\citep{frame}.

In the mission scenario detailed in Section \ref{sec:missionplan}, two spacecraft will form a  $\sim$90$^\circ$ angle for most of the cruise time (see Fig. \ref{fig:SS}). While our sensitivity calculations are for a single spacecraft, adding a second independent tracking system would allow to reduce the noise via cross-correlation of the signals, and as shown in Section \ref{sec:local}, would help with triangulating the source on the sky. Furthermore, if the signal analysis issues with low frequencies mentioned in \citet{godtierpaper} were resolved, we would expect  significantly more detections around the $10^{-5} - 10^{-4}$~Hz range, where the detection rate per frequency bin peaks. With a significant improvement in the total Allan deviation, a Doppler tracking experiment might become as capable as LISA at such low frequencies, and help bridge the gap between mHz detectors and PTAs. 
All of this, combined with the cost efficiency of a Doppler tracking experiment,  future ice giant missions would be a unique and cheap opportunity for low-frequency GW astronomy.

\vspace{-0.4cm}
\section*{Acknowledgements}
LZ acknowledges the support from the Swiss National Science Foundation under the Grant 200020\_178949.
We thank the referee Neil Cornish, and Leonid Gurvits, Sergei Pogrebenko, Lucio Mayer, Ravit Helled and Pedro R. Capelo for their useful comments.
DS is grateful for the short but meaningful time with \c{S}ansl{\i}.
\vspace{-0.45cm}
\section*{Data Availability}
The JPL HORIZONS System is publicly accessible.

\vspace{-0.2cm}
\bibliographystyle{mnras}
\bibliography{example} 





\bsp	
\label{lastpage}
\end{document}